\documentclass[useAMS,usenatbib]{mn2e}

\def\dg{$^{\circ}$}

\usepackage{graphicx}  
 \usepackage{times}
  \usepackage{lscape}
  \usepackage{rotating}
  \usepackage{txfonts}

\title[Structural changes in the hot Algol OGLE-LMC-DPV-097]{Structural changes in the hot Algol OGLE-LMC-DPV-097 and its disk related to its long-cycle}
\author[Garc\'es L., Mennickent, R.E.,  Djura{\v s}evi{\'c}, G. et al. ]
  {J. Garc\'es L.$^{1}$\thanks{E-mail: jgarcesletelier@gmail.com},  
  R.E.  Mennickent$^{1}$,
  Djura{\v s}evi{\'c}$^{2,3}$,
  R. Poleski $^{4,5}$,
     I. Soszy{\'n}ski$^{4}$     \\
  $^1$Universidad de Concepci\'on, Departamento de Astronom\'{\i}a,
      Casilla 160-C, Concepci\'on, Chile\\
       $^{2}$ Astronomical Observatory, Volgina 7, 11060 Belgrade 38, Serbia   \\
 $^{3}$ Isaac Newton Institute of Chile, Yugoslavia Branch\\
          $^{4}$ Warsaw University Observatory, Al. Ujazdowskie 4, 00-478 Warszawa, Poland \\ 
          $^{5}$ Department of Astronomy, Ohio State University, 140 W. 18th Ave., 
Columbus, OH 43210, USA\\
       }
\date{}

\begin{document}


\maketitle 

\begin{abstract} 
Double Periodic Variables (DPVs) are hot Algols showing a long photometric cycle of uncertain origin. 
We report the discovery of changes in the orbital light curve of OGLE-LMC-DPV-097 which depend on the phase of its long photometric cycle.
During the ascending branch of the long-cycle the brightness at the first  quadrature is larger than during the second  quadrature, 
during the maximum of the long-cycle the brightness is basically the same at both  quadratures, 
during the descending branch the brightness at the second  quadrature is larger than during
the first  quadrature and during the minimum of the long-cycle the secondary minimum disappears. 
We model the light curve at different phases of the long-cycle and find that the data are consistent with changes in the properties of the accretion  disk and two  disk spots.  The  disk's size and temperature change with the long-cycle period.
We find a smaller and hotter  disk at minimum and larger and cooler  disk at maximum. 
The spot temperatures, locations and angular sizes also show variability during the long-cycle. 
  
\end{abstract}


\section{Introduction}

In the close interacting binaries Double Periodic Variables (DPVs) a giant fills its Roche lobe and transfers mass  on to a B-type 
dwarf feeding an optically thick accretion  disk. About 250 DPVs have been found in the Galaxy and the Magellanic Clouds
\citep{2010AcA....60..179P, 2013AcA....63..323P, 2017SerAJ.194....1M}. Typical orbital periods are few days and long-cycles last hundreds of days. The long-cycle lasts in average about 33 times the orbital period, but single period ratios run typically between 27 and 39.
DPVs have been recently reviewed by \citet{2016MNRAS.455.1728M} and  \citet{2017SerAJ.194....1M}.

The more enigmatic fingerprint of a DPVs is its long-cycle. It has been attributed to  recurrent mass loss 
\citep{2008MNRAS.389.1605M, 2012MNRAS.427..607M}. 
 The triggering mechanism might be the magnetic dynamo of the rapidly rotating donor, that should modulate the mass transfer through the inner Lagrangian point  \citep{2017A&A...602A.109S}. A careful inspection of these  hypotheses is needed in a number of systems before to validate them.

In this paper we present our analysis of the light curve of  OGLE-LMC-DPV-097 ($\alpha_{2000}$= 05:34:18.06, $\delta_{2000}$= -70:38:08.7; MACHO 11.8746.65, 2MASS J05341808-7038087). This object  
is classified as a Double Periodic Variable with orbital period 7\fd751749 $\pm$ 0\fd000202 and long-cycle  302\fd622 $\pm$  0\fd109 by \citet{2010AcA....60..179P}.  
OGLE-LMC-DPV-097   has one of the largest amplitude  long-cycles among DPVs,
viz.\, 0.717 mag in the $I$ band. In this paper we extend the data analyzed by \citet{2010AcA....60..179P} including OGLE-IV photometry and present new
and interesting results potentially useful to explain the DPV phenomenon.

\section{Photometric data}

The photometric time-series analyzed in this paper are taken from the 
OGLE-III/IV project databases\footnote{OGLE-III/IV data kindly provided by the OGLE team.}. The OGLE-IV project is described by \citet{2015AcA....65....1U}.
The whole dataset, summarized in Table 1, spans a time interval of 4569 days, i.e. 12.5 years.


\begin{table}
\centering
 \caption{Summary of survey photometric observations. The number of measurements, starting and ending times for the series and average magnitude are given. HJD zero point is 2\,450\,000.
 Single point uncertainties are between 4 and 6 mmag.}
 \begin{tabular}{@{}lcrrcc@{}}
 \hline
Database &N &$HJD_{\rm start}$ &$HJD_{\rm end}$&mag &band \\
\hline
III  & 443 & 2167.89750 & 4954.53850 & 16.086 & $I$ \\
IV   & 580 & 5260.63876 & 6736.62011 & 16.140 & $I$ \\
 \hline
\end{tabular}
\end{table}

\section{Results}

 \subsection{Light curve disentangling}
 
 We disentangled the light curve into an orbital and long-cycle part with the aid of a Fourier decomposition algorithm
 described  by \citet{2012MNRAS.421..862M} and the ephemerides for the orbital and long periods given by \citet{2010AcA....60..179P}.
 The orbital and long-cycle light curves phased with the respective periods show  an interesting 
 behaviour  that can be summarized as (Fig.\,1): (a) during the ascending branch of the long-cycle the brightness at the first  quadrature ($\Phi_{\rm o}$ = 0.25) is larger than during the second  quadrature ($\Phi_{\rm o}$ = 0.75), (b) 
during the maximum of the long-cycle the brightness is basically the same at both  quadratures, 
(c) during the descending branch the brightness at the second  quadrature ($\Phi_{\rm o}$ = 0.75) is larger than during
the first  quadrature ($\Phi_{\rm o}$ = 0.25) and (d) during the minimum of the long-cycle the secondary minimum disappears. 
Furthermore, we have found that the variability is larger around the eclipses and it occurs smoothly during the long-cycle (Fig.\,2)

The above  behaviour suggests that there are structural changes in the binary (stars, gas stream,  disk, circumstellar matter) related to the
long-cycle. This is a strong  constraint for any competent model intending to explain the long-cycle in this DPV.

\begin{figure}
\scalebox{1}[1]{\includegraphics[angle=0,width=8cm]{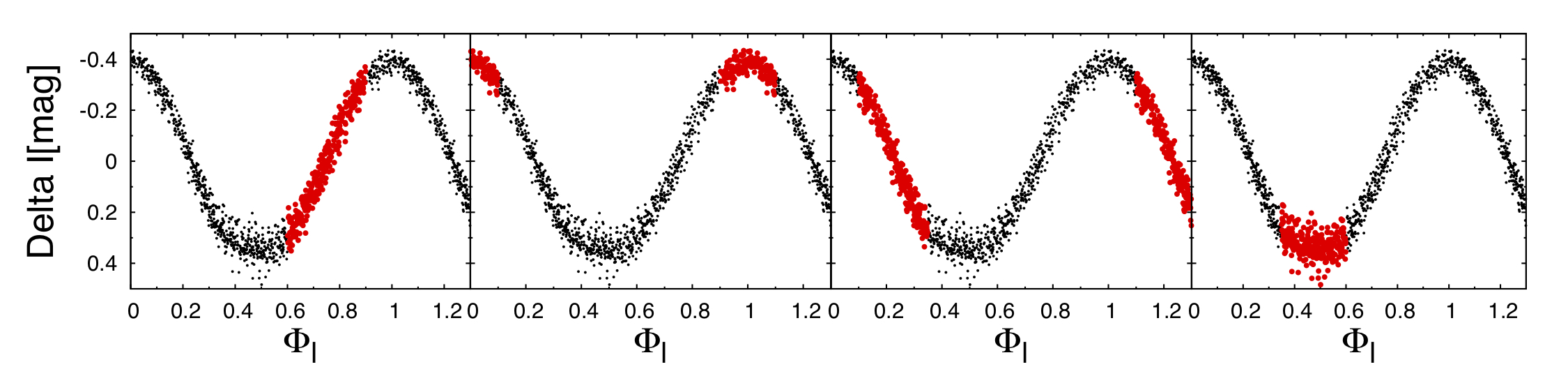}}\\
\scalebox{1}[1]{\includegraphics[angle=0,width=8cm]{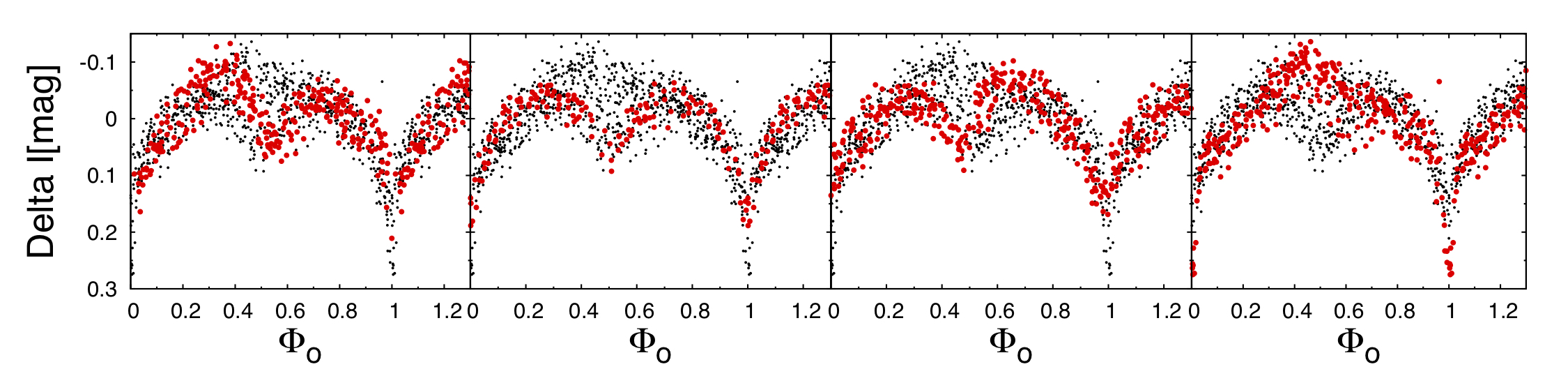}}
\caption{ Disentangled long-cycle (up) and orbital (down) light curves phased with the respective periods.
Black dots show the complete dataset, red dots show segments of the data  of the long-cycle. It is evident the change
in orbital light curve shape at different long-cycle phases.
   }
  \label{x}
\end{figure}

\begin{figure}
\scalebox{1}[1]{\includegraphics[angle=0,width=8cm]{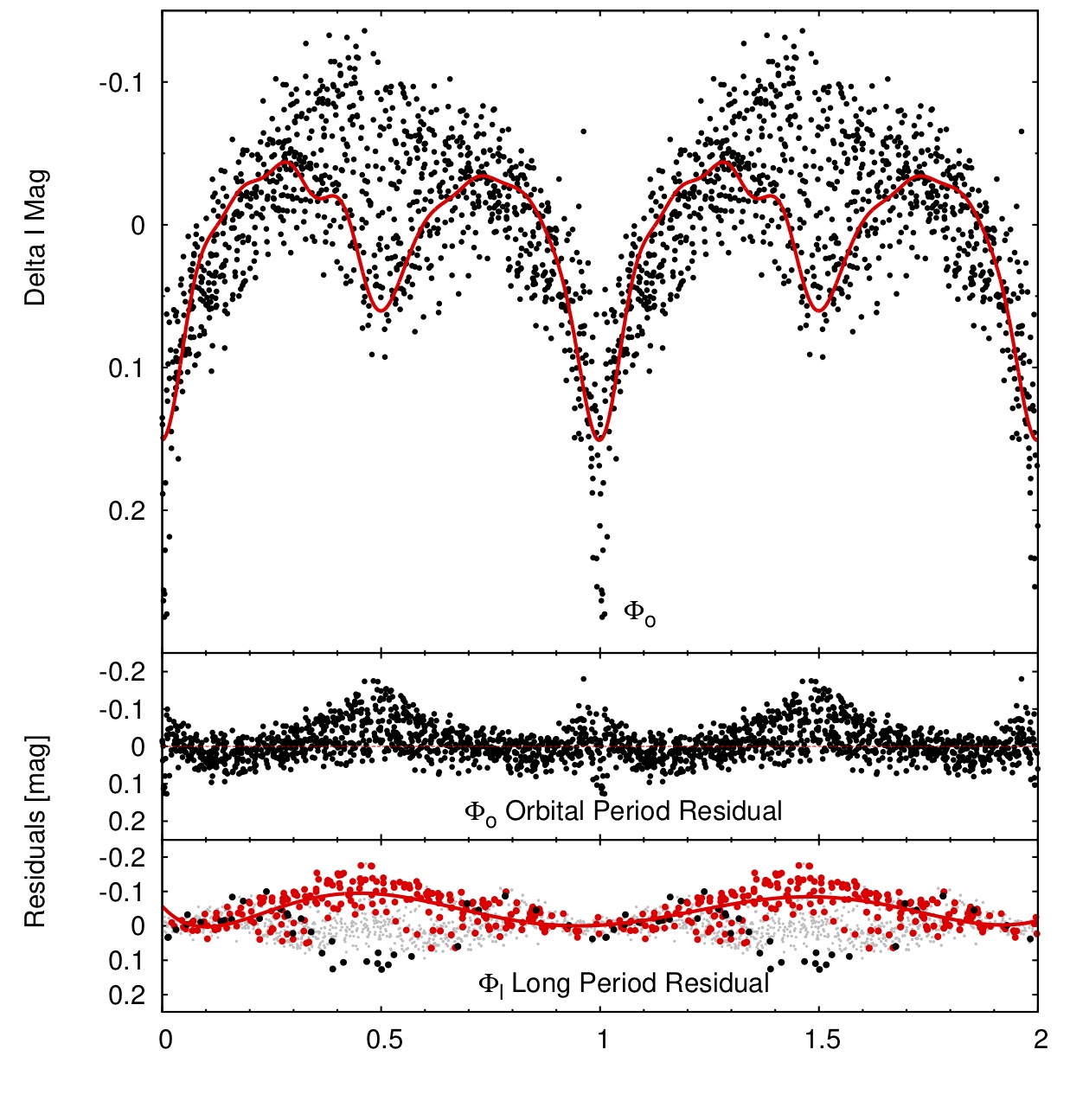}}\\
\caption{The upper panel shows the orbital light curve and a fit to the data obtained on maximum (red line).
The middle panel shows the residuals of the fit, indicating that the larger variability occurs on times of eclipses. 
The bottom panel shows the same residuals that above, but in the x-axis the long-cycle phase is shown. Red points are data taken around the secondary eclipse (0.45 $\leq \Phi_{\rm o} \leq$ 0.55)  and black points around the main eclipse (0.98 $\leq \Phi_{\rm o} \leq$ 1.02); 
the changes in the eclipses depth occur smoothly during the long-cycle. 
   }
  \label{x}
\end{figure}

 \subsection{Light curve model}
 
In order to qualitatively understand the orbital light changes we separately analyzed the orbital light curves on the ascending, top, descending and minimum branches with a program developed by \citet{1992Ap&SS.196..267D} which implements a sophisticated and versatile binary star model based on Roche geometry. This program has been used and tested for more than two decades on a wide range of binary configurations  
\citep[e.g.][]{2010MNRAS.409..329D, 2015MNRAS.448.1137M}. The system parameters that best fit the observed light curve are estimated using the Marquart-Levenberg algorithm (Marquardt 1963) with modifications described  in \citet{1992Ap&SS.197...17D} to minimize the sum of squared residuals between the observed (O) and calculated (C)  light curves.

In absence of spectroscopic data we use educated approximations to model the system.
OGLE-LMC-DPV-097 has MACHO catalogue\footnote{http://vizier.u-strasbg.fr/viz-bin/VizieR}  colour $V-R$ = -0.046 so we use $T_1$ = 14\,kK for the gainer. In addition, 
we assume synchronous rotation for the stellar components and $q$ = 0.2.
Based on the studied systems these figures can be considered representative for a DPV \citep{2016MNRAS.455.1728M}.  
With this approximation, we don't intend to get accurate system parameters for our system, at least not in this study, but
 to get a qualitative description of the {\it differential changes} occurring in the system during the long-cycle. For this reason we don't
focus on the absolute stellar properties in this paper. A trial using $q$ = 0.15 gives similar qualitative results  to reported in this paper and 
suggests typical errors of 20  per cent for the  disk radius, temperature and spots temperature ratio.

Following earlier work with DPVs, our model assumes an optically and geometrically thick  disk. The  disk edge is approximated by a cylindrical surface.The thickness of the  disk can change linearly with radial distance, allowing the  disk to take a conical shape (convex, concave or plane-parallel). The geometrical properties of the  disk are determined by its radius ($R_{\rm d}$), its thickness at the edge ($d_{\rm e}$) and the thickness at the  centre ($d_{\rm c}$). The cylindrical edge of the  disk is characterized by its temperature, $T_{\rm d}$, and the conical surface of the  disk by a radial temperature profile that follows the temperature distribution proposed by Zola (1991):\\

$T (r) = T_{\rm d} + (T_{\rm h} - T_{\rm d}) [1 - (\frac{r - R_{\rm h}}{ R_{\rm d} - R_{\rm h}})]^{a_{\rm T}}$ \hfill(1) \\

We assume that the  disk is in physical and thermal contact with the gainer, so the inner radius and temperature of the  disk are equal to the temperature and radius of the star. The temperature of the  disk at the edge ($T_{\rm d}$) and the temperature exponent ($a_{\rm T}$ ), as well as the radii of the star ($R_{\rm h}$) and of the  disk ($R_{\rm d}$) are free parameters, determined by solving the inverse problem. We refined 
the model of the system by introducing hot active regions on the edge of the  disk. 
Our model includes two such active regions: a hot spot (hs) and a bright spot (bs). These regions are characterized by their temperatures, angular dimensions and longitudes.  These parameters are also determined by solving the inverse problem and are given in Table\,2.  The models are compared with the observations in Figs. 3 and 4.

\section{Discusion}

We find that changes in orbital light curve shape can be explained as changes in system parameters, especially those related to the
 disk and hot/bright spots. 
The  disk is smaller and hotter at long-cycle minimum  while it is  
larger and cooler at long-cycle maximum. The amplitude of the temperature variation is 2840\,K, being the 
minimum reached during the ascending branch. 
The hot spot has maximum temperature relative to the  disk at minimum of the long-cycle while the bright spot does it 
during the ascending branch. While the hotspot
  position shows a small variation during the long-cycle with  $\lambda_{\rm h s}$ = 341\fdg5 -- 349\fdg2, 
 the bright spot moves along the disk outer rim during the long-cycle with  $\lambda_{\rm bs}$ = 114\fdg6 -- 255\fdg2. 
We notice that the hotspot in 7 DPVs is found at $\lambda_{\rm hs}$ (average) = 327\fdg7 $\pm$ 5\fdg6 (std) while the bright spot in the same  systems 
is found in the range $\lambda_{\rm bs}$ = 107\dg -- 162\dg \citep{2017SerAJ.194....1M}. This suggests that bright spots tends to occupy broader  disk regions than hotspots. In LMC-DPV-97, when the bright spot 
 is opposite to the donor, the secondary eclipse disappears, this happens at long-cycle minimum coinciding with the smaller  disk size.
 
Moving bright spots have been reproduced by 3D simulations of gas dynamics in the cataclysmic variable V\,455\,And, and interpreted  
in terms of interactions of a density wave (DW) with shock regions; in this model the DW  is in retrograde precession with angular velocity few percent lower than the orbital velocity of the
system. As a result, ``the times when the density wave interacts with the shocks will be increasing delayed in orbital phase with each successive rotation" \citep{2015ARep...59..191K}. Future studies could indicate if this model can be  successful for explaining the moving bright spots found in LMC-DPV-97.

We notice that the increase in hotspot angle $\lambda_{\rm hs}$ is in principle  consistent with the increase in disc size, if
the hotspot is the site where a free-fall stream hits the disk  \citep{1975MNRAS.170..325F}. Even if the stream interacts with a gas envelope
forming a hotline in the outer disk border as suggested by  \citet{1997ARep...41..786B}, the spread of this region along with the low mass ratio might help to keep small the
variability of the  $\lambda_{\rm hs}$ parameter.

\begin{table}

\caption{Results of the analysis of the orbital light curves
obtained by solving the inverse problem for the Roche model with
an accretion disk around the more-massive (hotter) component.}
\label{DPV LMC}
\[
\begin{array}{@{\extracolsep{+0.0mm}}lrrrr@{}}
\hline
\noalign{\smallskip}
{\rm Quantity} &{\rm asc} & {\rm max} & {\rm des} & {\rm min}  \\
\noalign{\smallskip}
\hline
\noalign{\smallskip}
n 					        & 370	& 200          & 380   		 & 387           \\
{\rm \Sigma(O-C)^2}			& 0.3310	& 0.0772           & 0.2439		& 0.2754 	\\
{\rm \sigma_{rms}}			& 0.0299	& 0.0197           & 0.0254		& 0.0267 	\\
i {\rm [^{\circ}]}                           & 74.3       & 75.1                & 74.4       & 74.4          \\
{\rm F_d}				        &  0.93 	& 0.94             & 0.68   		& 0.46   	\\
{\rm T_d} [{\rm K}]			&  4030 	&  5580            &  5210  		& 6870  	\\
{\rm d_e} [a_{\rm orb}]	        & 0.044 	& 0.034            & 0.082  		& 0.031  	\\
{\rm d_c} [a_{\rm orb}]		& 0.100 	& 0.110            & 0.102  		& 0.084  	\\
{\rm a_T}				         & 4.3	& 6.5                  & 5.9   	  & 5.1   	\\
{\rm F_h}				         & 0.242 	& 0.242            & 0.242  		& 0.242  	\\
{\rm T_h} [{\rm K}]			& 14000 	& 14000             & 14000  		& 1400  	\\
{\rm T_c} [{\rm K}]			&  4930 	&  4980              &  4910  		&  4950  	\\
{\rm A_{hs}=T_{hs}/T_d}		&  1.63	&  1.70            &  1.14  		&  1.71 	\\
{\rm \theta_{hs}}{\rm [^{\circ}]}	&  19.4 	&  19.1            &  14.8  		&  23.0 	\\
{\rm \lambda_{hs}}{\rm [^{\circ}]} & 348.6 	& 349.2            & 340.5  		& 341.6 	\\
{\rm \theta_{rad}}{\rm [^{\circ}]} 	 & -1.0  	 & -3.6             &  8.3       	&  2.4  	\\
{\rm A_{bs1}=T_{bs}/T_d}		 &  1.78 	&  1.17             &  1.16  		&  1.76 	\\
{\rm \theta_{bs}} {\rm [^{\circ}]}	 &  23.3 	&  54.4             &  26.2  		&  49.0 	\\
{\rm \lambda_{bs}}{\rm [^{\circ}]} & 137.2 	& 114.6               & 255.2 	 & 169.6 	\\
{\rm \Omega_h}			 & 8.682 	& 8.699            & 8.681 		& 8.681 	\\
{\rm \Omega_c}				 & 2.233 	& 2.233             & 2.233  		& 2.233 	\\
\noalign{\smallskip} \hline \noalign{\smallskip}
\cal M_{\rm_h} {[\cal M_{\odot}]}       & 5.51  	& 5.51  & 5.51  		& 5.51  	\\
\cal M_{\rm_c} {[\cal M_{\odot}]}	       & 1.10  	& 1.10  & 1.10  		& 1.10  	\\
\cal R_{\rm_h} {\rm [R_{\odot}]}	       & 3.65  	& 3.64  & 3.65  		& 3.65  	\\
\cal R_{\rm_c} {\rm [R_{\odot}]}	       & 7.75        & 7.75  & 7.75  		& 7.75  	\\
{\rm log} \ g_{\rm_h}			       & 4.06  	& 4.06  & 4.06  		& 4.06  	\\
{\rm log} \ g_{\rm_c}			       & 2.70  	& 2.70  & 2.70  		& 2.70  	\\
M^{\rm h}_{\rm bol}			        &-1.87  	&-1.86  &-1.87  		&-1.87  	\\
M^{\rm c}_{\rm bol}			        & 1.03  	& 0.99  & 1.04  		& 1.01  	\\
a_{\rm orb}  {\rm [R_{\odot}]}		& 30.91 	& 30.91 & 30.91 		& 30.91 	\\
\cal{R}_{\rm d} {\rm [R_{\odot}]}       	& 15.28 	& 15.40  & 11.20 		&  7.49 	\\
\rm{d_e}  {\rm [R_{\odot}]}		        &  1.36 	&  1.06 &  2.53 		&  0.95  	\\
\rm{d_c}  {\rm [R_{\odot}]}		        &  3.09 	&  3.41 &  3.16 		&  2.6 	        \\
\noalign{\smallskip}
\hline
\end{array}
\]
\smallskip

FIXED PARAMETERS: $q={\cal M}_{\rm c}/{\cal M}_{\rm h}=0.20$ - mass ratio of
the components, ${\rm T_h=14000 K}$  - temperature of the more-massive (hotter)
gainer,
${\rm F_c}=1.0$ - filling factor for the critical Roche lobe of the donor,
$f{\rm _h}=1.0 ; f{\rm _c}=1.00$ - non-synchronous rotation coefficients of
the gainer and donor respectively, ${\rm \beta_h=0.25 ; \beta_c=0.08}$ -
gravity-darkening coefficients of the gainer and donor, ${\rm A_h=1.0 ;
A_c=0.5}$  - albedo coefficients of the gainer and donor.

\smallskip \noindent Quantities: $n$ - number of observations,
${\rm \Sigma (O-C)^2}$ - final sum of squares of residuals between observed
(LCO) and synthetic (LCC) light-curves, ${\rm \sigma_{rms}}$ - root-mean-square
of the residuals, $i$ - orbit inclination (in arc degrees),
${\rm F_d=R_d/R_{yc}}$ - disk dimension factor (ratio of the disk radius to
the critical Roche lobe radius along y-axis), ${\rm T_d}$ - disk-edge
temperature, $\rm{d_e}$, $\rm{d_c}$,  - disk thicknesses (at the edge and at
the  centre of the disk, respectively) in the units of the distance between the
components, $a_{\rm T}$ - disk temperature distribution coefficient,
${\rm F_h=R_h/R_{zc}}$ - filling factor for the critical Roche lobe of the
hotter, more-massive gainer (ratio of the stellar polar radius to the
critical Roche lobe radius along z-axis), 
${\rm T_c}$ - temperature of the less-massive (cooler) donor,
${\rm A_{hs,bs}=T_{hs,bs}/T_d}$ - hot and bright spots' temperature
coefficients, ${\rm \theta_{hs,bs}}$ and ${\rm \lambda_{hs,bs}}$ -
spots' angular dimensions and longitudes (in arc degrees), 
${\rm \theta_{rad}}$- angle between the line perpendicular to the local disk edge surface and the
direction of the hot-spot maximum radiation, ${\rm \Omega_{h,c}}$ -
dimensionless surface potentials of the hotter gainer and cooler donor,
$\cal M_{\rm_{h,c}} {[\cal M_{\odot}]}$, $\cal R_{\rm_{h,c}} {\rm [R_{\odot}]}$
- stellar masses and mean radii of stars in solar units,
${\rm log} \ g_{\rm_{h,c}}$ - logarithm (base 10) of the system components
effective gravity, $M^{\rm {h,c}}_{\rm bol}$ - absolute stellar bolometric
$a_{\rm orb}$ ${\rm [R_{\odot}]}$, $\cal{R}_{\rm d} {\rm [R_{\odot}]}$,
$\rm{d_e} {\rm [R_{\odot}]}$, $\rm{d_c} {\rm [R_{\odot}]}$ - orbital semi-major axis,
disk radius and disk thicknesses at its edge and  centre, respectively, given in the
solar radius units.
\end{table}

 \section{Conclusions}

We find changes in the orbital light curve of DPV097 linked to the phase of the long-cycle. 
A preliminary analysis of the light curve indicates that changes in  disk size/temperature and spot temperature/position
can explain the different observed light curves during the long-cycle. Especially interesting 
is the result that at minimum of the long-cycle the  disk is smaller and hotter and the spots are also hotter, while at maximum
the  disk is larger and cooler. A more detailed study of this object is in preparation.

\begin{figure}
\scalebox{1}[1]{\includegraphics[angle=0,width=8cm]{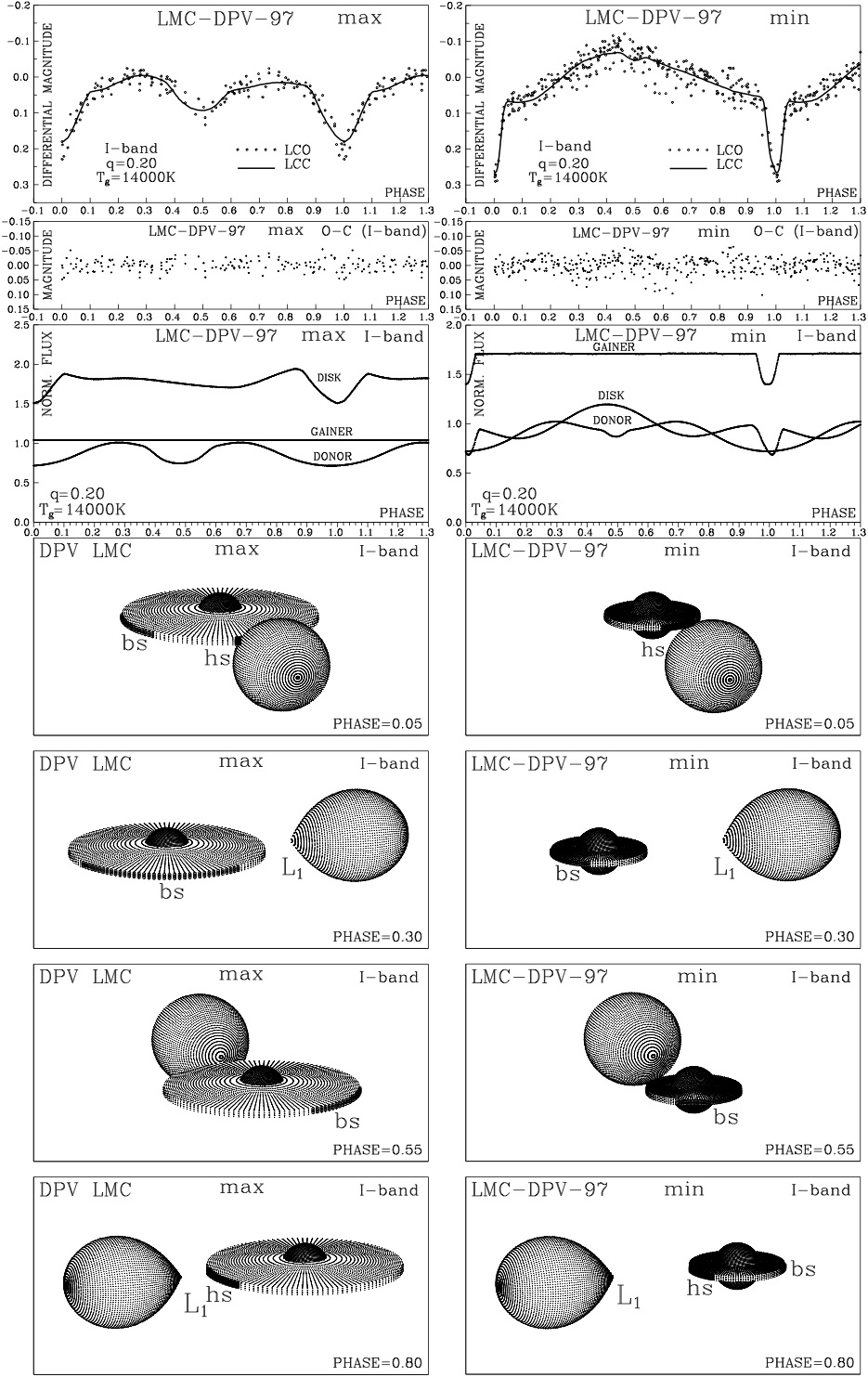}}
\caption{The model compared with observations on the maximum and minimum  of the long cycle. Relative flux contributions
are in the third panel and representative views of the system at different phases are also given.}
  \label{x}
\end{figure}

\begin{figure}
\scalebox{1}[1]{\includegraphics[angle=0,width=8cm]{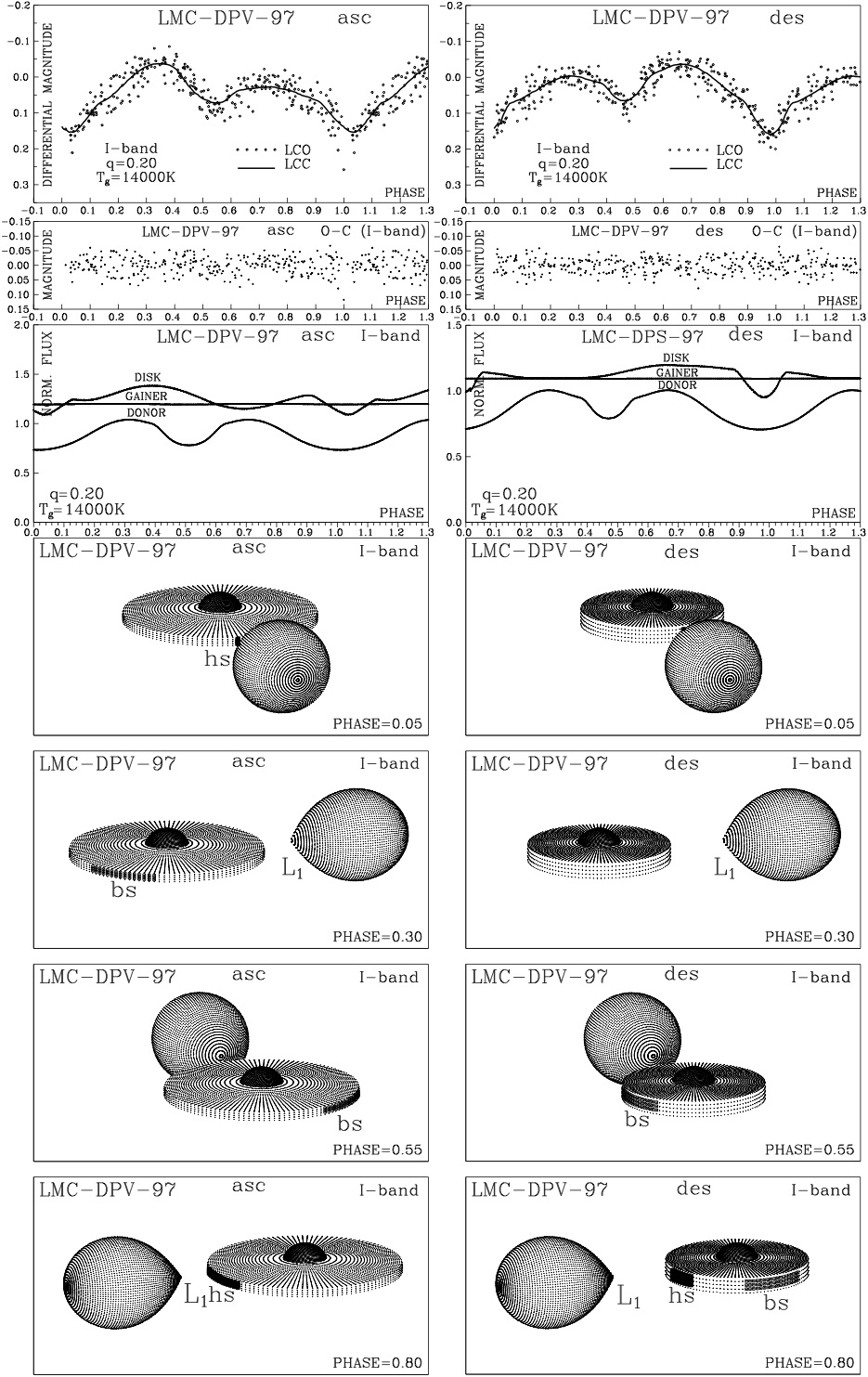}}
\caption{The model compared with observations and residuals on the ascending and descending branches of the long cycle.
Relative flux contributions
are in the third panel and representative views of the system at different phases are also given.}
  \label{x}
\end{figure}

\section{Acknowledgments}

 Thanks to the anonymous referee for providing useful comments that improved the first version of this manuscript.
This research has made use of the SIMBAD database, operated at CDS, Strasbourg, France.
R.E.M. acknowledges support by VRID-Enlace 216.016.002-1.0 and the BASAL Centro de Astrof{\'{i}}sica y Tecnolog{\'{i}}as Afines (CATA) PFB--06/2007. 
The OGLE project has received funding from the Polish National Science
Centre grant MAESTRO no. 2014/14/A/ST9/00121.

\bsp 
\label{lastpage}

\newpage

\end{document}